\documentclass[12pt]{article}
\usepackage{amsfonts, amssymb}

\newtheorem{theorem}{Theorem}[section]
\newtheorem{lemma}{Lemma}[section]
\newtheorem{corollary}{Corollary}[section]
\newtheorem{observation}{Observation}[section]

\begin{document}

\title{On Faces of the set of Quantum Channels}

\author{Raphael Loewy\\
Department of Mathematics\\
Technion -- Israel Institute of Technology\\
\tt{loewy@tx.technion.ac.il}}
\maketitle

\begin{abstract}

A linear map $L$ from ${\mathbb C}^{n \times n}$ into
${\mathbb C}^{n \times n}$ is called a quantum channel if it is
completely positive and trace preserving.
The set ${\cal L}_n$ of all such quantum channels is known to be a
compact convex set. While the extreme points of ${\cal L}_n$ can be
characterized, not much is known about the structure of its higher
dimensional faces. Using the so called Choi matrix
$Z(L)$ associated with the quantum channel $L$, we compute the maximum
dimension of a proper face of ${\cal L}_n$, and in addition the possible
dimensions of faces generated by $L$ when $rank \  Z(L)=2 $.

\end{abstract}



\newpage

\setcounter{section}{-1}
\section{Introduction}

Let ${\mathbb C}^{n \times n}$, ${\cal H}_n$ and $PSD_n \subset {\cal H}_n$
denote, respectively, the sets of all $n \times n$ complex matrices, complex hermitian matrices and positive semidefinite matrices.
A \emph{pure state} is a positive semidefinite matrix of rank $1$ and trace $1$. Let ${\cal P}_n$ denote the set of pure states in ${\cal H}_n$.
Let $[m]=\{1,2,\ldots , m\}$ for any positive integer $m$.

A linear operator $L:{\mathbb C}^{n \times n} \rightarrow {\mathbb C}^{n \times n} $ is called
\emph{completely positive} if
\begin{equation}
L(X) = \sum_{i=1}^{k}A_iXA_i^* \ , \
A_i \in {\mathbb C}^{n \times n} \ , \
i \in [k].
\label{EQ0.1}
\end{equation}
A completely positive operator which is also trace preserving is called \emph{quantum channel}. It is known and straightforward to see that if $L$ is given by
(\ref{EQ0.1}) then it is a quantum channel if and only if
\begin{equation}
\sum_{i=1}^{k}A_i^*A_i = I_n,
\label{EQ0.2}
\end{equation}
where $I_n$ denotes the identity matrix of order $n$.

Let ${\cal L}_n$ denote the set of all quantum channels defined on
${\mathbb C}^{n \times n}$. Then ${\cal L}_n$ is a compact convex set, and it is the purpose of this paper to consider its face structure.
The $0$-dimensional faces are the extreme points of ${\cal L}_n$, and they are of fundamental importance, thus attracting significant attention, cf.
\cite{Ch}, \cite{FL}, \cite{RSW}, \cite{Ru}.
One characterization, due to Choi \cite{Ch}, is given in terms of the matrices $A_i$ appearing in (\ref{EQ0.1}).
Another way to decide whether $L \in {\cal L}_n$ is an extreme point is to use the so called \emph{Choi matrix} $Z(L)$ associated with $L$.

For $i,j \in [n]$ let $E_{ij} \in {\mathbb C}^{n \times n}$
denote the matrix with $1$ in the $ij$ position and $0$'s elsewhere. Then
\begin{equation}
\label{EQ0.3}
Z(L) = \left[L(E_{ij})\right]_{i,j=1}^{n}.
\end{equation}

It is well known that a linear operator
$L: {\mathbb C}^{n \times n} \rightarrow {\mathbb C}^{n \times n}$
is completely positive if and only if $Z(L)$ is positive semidefinite.
Moreover, letting
$A=Z(L) =[A_{ij}]_{i,j=1}^{n}$
with $A_{ij} \in {\mathbb C}^{n \times n}$ for $i,j \in [n]$,
then a completely positive $L$ is a quantum channel if and only if the
trace conditions $tr A_{ij}=\delta_{ij}$ also hold for $i,j \in [n]$.

Friedland and Loewy \cite{FL} gave a necessary and sufficient condition
for $L \in {\cal L}_n$ to be an extreme point in terms of the null space of
$Z(L)$.
It is pointed out in \cite[Remark 6]{Ch}, see also Ruskai, Szarek Werner
\cite{RSW} and Ruskai \cite{Ru}, that if $L \in {\cal L}_n$ is an extreme point, then $rank Z(L) \leq n$.

It seems that higher dimensional faces of ${\cal L}_n$ have not attracted
much attention, although they are of interest as well as the extreme points.
Our purpose here is to consider some of those faces. The paper is organized as
follows:
In Section 1 we give some additional notation and some preliminary results. In Section 2 we consider faces generated by matrices $Z(L)$ such that
$rank Z(L) =2$ and determine their possible dimensions. In Section 3 we show the maximum dimension of a proper face of ${\cal L}_n$ is $n^4-3n^2+1$.
The concept of \emph{face} is basic in convex set theory, and the reader is refered to \cite{Ro} for related information.

\section{Preliminaries}
\setcounter{equation}{0}

We first introduce some additonal notation.
Given $A \in {\mathbb C}^{n \times n}$
and $\alpha , \beta \subset [n]$ let
$A[\alpha | \beta]$ be the submatrix of $A$ based on rows in $\alpha$ and
columns in $\beta$, and let
$A[\alpha ] = A[\alpha | \alpha]$.
Also, given any matrix $A$ denote by $A^{(j)}$ its j-th column.
Given positive integers $r$ and $s$ let ${\cal J}_{rs}$  be the all ones matrix of order $r \times s$, and let ${\cal J}_r = {\cal J}_{rr}$.
We use the standard inner product, denoted by $\langle \cdot , \cdot \rangle$.

Let
\begin{equation}\label{EQ1.1}
{\cal C}_n = \left\{ A = [A_{ij}]_{i,j=1}^{n} \in PSD_{n^2} \ , \
A_{ij} \in {\mathbb C}^{n \times n} \ , \
tr A_{ij} = \delta_{ij} \ , \ i,j \in [n]\right\}.
\end{equation}
Then, ${\cal C}_n$ is a compact convex set in the real vector space
${\cal H}_{n^2}$ of dimension $n^4-n^2$, and the map from ${\cal L}_n$ to
${\cal C}_n$ defined by $L \rightarrow Z(L)$
is an isomorphism (see \cite{FL}).
Given any $A \in {\cal C}_n$, denote  by ${\cal F}(A)$
the face of ${\cal C}_n$ generated by $A$.

The smallest affine subspace of ${\cal H}_{n^2}$ containing ${\cal C}_n$ is
\begin{equation}
\label{EQ1.2}
{\cal W}_n = \left\{ A = [A_{ij}]_{i,j=1}^{n} \in {\cal H}_{n^2} \ , \
A_{ij}\in {\mathbb C}^{n \times n} \ , \
tr A = \delta_{ij} \ , \ i,j \in[n] \right\}.
\end{equation}

We note that the (relative) interior of ${\cal C}_n$ consists of all
positive definite matrices in ${\cal C}_n$.
Indeed, every positive definite matrix in ${\cal C}_n$ is in its
(relative) interior. Since $\frac{1}{n}I_{n^2} \in {\cal C}_n$,
it follows that if $A \in {\cal C}_n$ is singular then for
any $\varepsilon > 0$  \ \
$(1+\varepsilon) A - \frac{\varepsilon}{n} I_{n^2} \in {\cal W}_n$
and has a negative eigenvalue. Hence $A$ belongs to the (relative)
boundary of ${\cal C}_n$.

The following observation characterizes the rank 1 matrices in
${\cal C}_n$.

\begin{observation}
\label{observe1.1}
Let $x=(x_1^t,x_2^t, \ldots , x_n^t)^t \in {\mathbb C}^{n^2}$,
where $x_i \in {\mathbb C}^n$ for all $i \in [n]$.
Then $xx^*\in {\cal C}_n$ if and only if
$\{x_i\}_{i \in [n]}$ forms an orthonormal basis of
${\mathbb C}^n$ with respect to the standard inner product.
\end{observation}

\noindent {\bf Proof:} It is clear that $\{x_i\}_{i \in [n]}$
forms an arthonormal basis with respect to the standard inner product if and only if the trace conditions in (\ref{EQ1.1}) are satisfied by $xx^*$.\hspace{\fill} $\Box$

\vspace{10pt}

The next lemma is useful when considering faces of ${\cal C}_n$.
\begin{lemma}
\label{lemma1.1}
Let $A \in {\cal C}_n$ and $B \in {\cal F}(A)$. Then,
\begin{itemize}
\item[(a)] $\ker  A \subset \ker  B$,
\item[(b)] If $B$ belongs to the (relative) boundary of ${\cal F}(A)$, $rank B < rank A$.
\end{itemize}
\end{lemma}

\noindent {\bf Proof:} Since $A$ belongs to the (relative) interior of
${\cal F}(A)$ there exist\\ $0 < \alpha \leq 1$ and
$C \in {\cal C}_n$ such that $A = \alpha B +(1-\alpha)C$.
Then (a) follows because if $AX=0$ then $x^*Ax=\alpha x^*Bx+(1-\alpha)x^*Cx=0$, implying $Bx=0$.

To prove (b) assume $B$ belongs to the (relative) boundary of ${\cal F}(A)$.
For any $\varepsilon \in {\mathbb R}$,  $(1+\varepsilon)B -\varepsilon A \in {\cal W}_n$.
Suppose that $rank B = rank A$.
Then there exists $\varepsilon > 0$ sufficiently small such that
$(1+ \varepsilon)B -\varepsilon A$ is positive semidefinite, so belongs to
${\cal C}_n$.

This can be seen, for example, by the known result that $A$ and $B$
can be simultaneously diagonalized by a congruence,
and with the positive elements on both diagonal matrices appearing in the same
locations.
It follows that there exist $C \in {\cal C}_n$ and $0 < \alpha <1$
such that $B = \alpha A +(1-\alpha )C$,
so $B$ belongs to the (relative) interior of ${\cal C}_n$, a contradiction. \hspace{\fill} $\Box$

\vspace{10pt}

As a consequence of the previous lemma we get a simple geometric proof of the following known result.

\begin{corollary}
Every matrix in ${\cal C}_n$ is a convex combination of at most $n^2$ extreme points.
\end{corollary}

\noindent {\bf Proof:} Let $A \in {\cal C}_n$ and consider ${\cal F}(A)$.
It is well known that every extreme point of ${\cal F}(A)$ is also an
extreme point of ${\cal C}_n$.
Take such an extreme point, say $B$, and continue the line segment
from $A$ to $B$ until it meets the (relative) boundary
of ${\cal F}(A)$ at a point $C$. This is possible because ${\cal C}_n$ is compact.
Then $A$ is a convex combiantion of $B$ and $C$, and by Lemma \ref{lemma1.1}
$rank C < rank A$. Repeat the process with $C$. As $rank A \leq n^2$ the process consists of at most $n^2$ steps, completing the proof.
\hspace{\fill}
$\Box$

\section{The dimension of ${\cal F}(L)$ when $rank Z(L) =2$}
\setcounter{equation}{0}

Our goal here is to find out the possible dimensions that $F(L)$ can attain when $rank Z(L) =2$.

\begin{theorem}
Let $L \in {\cal L }_n$ with $rank Z(L) =2$. Then
\begin{itemize}
\item[(i)] $dim {\cal F}(L)$ is 0 or 2 if $n=2$,
\item[(ii)] $dim {\cal F}(L)$ is 0 or 1 or 2 if $n \geq 3$.
\end{itemize}
\end{theorem}

\noindent {\bf Proof:} It follows from Theorems 1 and 12 of \cite{FL} that
$L({\cal P}_m)$ contains a pure state, and since we can apply (independently)
unitary similarities in the domain and range, we can assume without loss of
generality that $L(E_{11}) =E_{11}$.
Also, for convenience, we write $A=Z(L)$.

\vspace{10pt}

\noindent (i) Suppose first that $n=2$. Then, as $L \in {\cal L}_2$,
we have (cf. (6.1) of \cite{FL})
\[
A =\left[\begin{array}{cccc}
1&0&0&y\\
0&0&0&0\\
0&0&1-c&s\\
\bar{y}&0&\bar{s}&c
\end{array}\right],
\]
and by assumption $rank(A)=2$.
As $A \in PSD_4$, we have $0 \leq c \leq 1$.
If $c=1$ then $s=0$ and $|y| < 1$. For every $B \in {\cal F}(A)$
$\ker B \supset \ker A$ by Lemma \ref{lemma1.1}, hence $e_2, e_3 \in \ker B$.
Therefore ${\cal F}(A)$ consists of all matrices
\[
\left[\begin{array}{cccc}
1&0&0&z\\
0&0&0&0\\
0&0&0&0\\
\bar{z}&0&0&1
\end{array}\right]
\]
with $z \in {\mathbb C}$, $|z| \leq 1$, so $dim {\cal F}(A)=2$.

\vspace{10pt}

Assume now that $0 \leq c < 1$. In this case the first and third columns of $A$ form a basis for its colmun space, and we must have
\begin{equation}
\label{EQ2.1}
\left[\begin{array}{c}
y\\
0\\
s\\
c
\end{array}\right] =
y\left[\begin{array}{c}
1\\
0\\
0\\
\bar{y}
\end{array}\right]
+ \frac{s}{1-c}\left[\begin{array}{c}
0\\
0\\
1-c\\
\bar{s}
\end{array}\right].
\end{equation}
Note the entries of $A$ have to satisfy
\begin{equation}
\label{EQ2.2}
(1-c)(c-|y|^2)=|s|^2.
\end{equation}
We will show now that if $B \in {\cal F}(A)$ then $B=A$,
implying that $dim {\cal F}(A)=0$.
It follows from (\ref{EQ2.1}) that $e_2$ and $e_4-ye_1-\frac{s}{1-c}e_3$
form a basis of $\ker A$,
and hence belong to $\ker B$ for $B \in {\cal F}(A)$. There exists $0 \leq c_1 \leq 1$
such that $b_{33}=1-c_1$. Hence $b_{14}=y$,
$b_{34}=\frac{s(1-c_1)}{1-c}$,
$b_{43}=\frac{\bar{s}(1-c_1)}{1-c}$
and
$b_{44}=yb_{41}+ \frac{s}{1-c}b_{43}=|y|^2+\frac{|s|^2(1-c_1)}{(1-c)^2}$.

\vspace{10pt}

As $b_{33}+b_{44}=1$ we get
\[
1-c_1+|y|^2+\frac{|s|^2(1-c_1)}{(1-c)^2}=1.
\]
Using (\ref{EQ2.2}), we get
\[
(1-c)^2(|y|^2-c_1)+(1-c)(c-|y|^2)(1-c_1)=0,
\]
so
\[
(c-c_1)(1-|y|^2)=0,
\]
and as $c<1$ implies $|y|<1$, we conclude that $c=c_1$,
$b_{33}=a_{33}$, $b_{34}=a_{34}$ and $b_{44}=a_{44}$. Hence $B=A$.

\vspace{10pt}

\noindent (ii) Suppose now that $n \geq 3$. The proof is by induction on $n$, although the induction hypothesis will not be used in all subcases to follow.

Recall that $L(E_{11})=E_{11}$, so rows and columns of $A$ indexed by
$2,3,\ldots,n$ are 0. This implies, as $L \in {\cal L}_n$, $a_{1,n+1}=0$.
We distinguish several cases.

\vspace{10pt}

\noindent (iia) Suppose that there exists a positive integer $i$,
$1 \neq i <n^2$, $i \equiv 1(\mbox{mod}\, n)$ such that $a_{ii}> 0$.
Without loss of generality we may assume

\begin{equation}
\label{EQ2.3}
a_{n+1,n+1}> 0.
\end{equation}
Since $rank A =2 = rank A[\{1,n+1\}]$,
we must have by a Schur complement argument
\begin{equation}
\label{EQ2.4}
A[\{n+2,n+3, \ldots , n^2\}]=
\end{equation}\[
A[\{n+2,n+3, \ldots , n^2\}|\{1,n+1\}]\left[\begin{array}{cc}
1&0\\0&a^{-1}_{n+1, n+1}
\end{array}\right]A[\{1,n+1\}|\{n+2,n+3, \ldots , n^2\}].
\]
In particular, (\ref{EQ2.4}) implies

\begin{equation}
\label{EQ2.5}
a_{n+j,n+j}=|a_{1,n+j}|^2+\frac{|a_{n+1,n+j}|^2}{a_{n+1,n+1}} \ ,\ j=2,3,\ldots , n.
\end{equation}
We must also have
\begin{equation}
\label{EQ2.6}
\sum_{j=n+1}^{2n}a_{jj}=1.
\end{equation}
We intend to show that $A$ is an extreme point in this case.
For that purpose, assume that
\[
A=\alpha C +(1-\alpha)D \ , \
C,D \in {\cal C}_n \ ,\
0 < \alpha < 1.
\]
By Lemma \ref{lemma1.1} $\ker C \supset \ker A$
and $\ker D \supset \ker A$.
The standard unit vectors $e_2,e_3, \ldots , e_n$ belong to $\ker A$,
so also to $\ker  C$, $\ker D$.
This means that rows and columns of $C$, as well as of $D$, indexed by $2,3,\ldots , n$ are 0. Also, trace conditions imposed by $C,D \in {\cal C}_n$ imply
\[
c_{11}=d_{11}=1 \ \mbox{and} \ c_{1i}=d_{1i}=c_{i1}=d_{i1}=0
\ \mbox{for} \ 1 \neq i < n^2 \ , \
i \equiv 1(\mbox{mod}\, n).
\]
The column vectors $A^{(1)}, A^{(n+1)} \in {\mathbb C}^{n^2}$ form a basis for the column space of $A$. So it is straightforward that
\begin{equation}
\label{EQ2.7}
A^{(j)}=a_{1j}A^{(1)}+
\frac{a_{n+1,j}}{a_{n+1,n+1}}A^{(n+1)} \ ,\ n+2 \leq j \leq n^2,
\end{equation}
implying that $e_j-a_{1j}e_1-\frac{a_{n+1,j}}{a_{n+1,n+1}}e_{n+1} \in \ker A
\subset \ker C, \ker D$
.

Hence, for $n+2 \leq j \leq n^2$,
\begin{equation}
\label{EQ2.8}
C^{(j)}=a_{1j}C^{(1)}+\frac{a_{n+1,j}}{a_{n+1,n+1}}C^{(n+1)} \ ; \
D^{(j)}=a_{1j}D^{(1)}+\frac{a_{n+1,j}}{a_{n+1,n+1}}D^{(n+1)}.
\end{equation}
This implies
\begin{equation}
\label{EQ2.9}
c_{1j}=a_{1j}c_{11}+\frac{a_{n+1,j}}{a_{n+1,n+1}}c_{n+1,1}=
1 \cdot a_{1j}+0\cdot \frac{a_{n+1,j}}{a_{n+1,n+1}}=a_{1j},
\end{equation}
and similarly
\begin{equation}
\label{EQ2.10}
d_{1j}=a_{1j}.
\end{equation}
Therefore,
\begin{equation}
\label{EQ2.11}
c_{j1}=d_{j1}=a_{j1}.
\end{equation}
Using $c_{n+1,1}=d_{n+1,1}=0$,
we obtain from (\ref{EQ2.8})
\begin{equation}
\label{EQ2.12}
c_{n+1,j}=\frac{a_{n+1,j}c_{n+1,n+1}}{a_{n+1,n+1}} \ ; \
d_{n+1,j}=\frac{a_{n+1,j}d_{n+1,n+1}}{a_{n+1,n+1}} \ ; \
\end{equation}
\[
c_{j, n+1} = \bar{c}_{n+1,j} \ ; \
d_{j, n+1} = \bar{d}_{n+1,j}.
\]

We let $n+2 \leq j \leq 2n$ and apply (\ref{EQ2.7}) and (\ref{EQ2.8}), using also
(\ref{EQ2.9}), (\ref{EQ2.10}), (\ref{EQ2.11}) and (\ref{EQ2.12}), to get
 \begin{equation}
\label{EQ2.13}
c_{jj}= |a_{ij}|^2+\frac{|a_{n+1,j}|^2c_{n+1,n+1}}{a^2_{n+1,n+1}},
\end{equation}
\begin{equation}
\label{EQ2.14}
d_{jj}= |a_{ij}|^2+\frac{|a_{n+1,j}|^2d_{n+1,n+1}}{a^2_{n+1,n+1}}.
\end{equation}
Substituting (\ref{EQ2.5}) into (\ref{EQ2.6}) yields
\[
a_{n+1,n+1} + \sum_{j=n+2}^{2n}\left(|a_{ij}|^2+ \frac{|a_{n+1,j}|^2}{a_{n+1,n+1}}\right)=1,
\]
so
\begin{equation}
\label{EQ2.15}
a_{n+1,n+1}^2+a_{n+1,n+1} \sum_{j=n+2}^{2n}|a_{ij}|^2 +
\sum_{j=n+2}^{2n}|a_{n+1,j}|^2 = a_{n+1,n+1}.
\end{equation}

Since $C \in {\cal C}_n$ we have
$\sum\limits_{j=n+1}^{2n}c_{jj}=1$,
so (\ref{EQ2.13}) implies
\begin{equation}
\label{EQ2.16}
a_{n+1,n+1}^2c_{n+1,n+1}+ a_{n+1,n+1}^2\sum_{j=n+2}^{2n}|a_{ij}|^2+
c_{n+1,n+1}\sum_{j=n+2}^{2n}|a_{n+1,j}|^2 = a_{n+1,n+1}^2.
\end{equation}
It follows from (\ref{EQ2.15}) and (\ref{EQ2.16}) that
\[
a_{n+1,n+1}^2\left(c_{n+1,n+1} - a_{n+1,n+1}\right)+
\left(c_{n+1,n+1} - a_{n+1,n+1}\right)
\sum_{j=n+2}^{2n}|a_{n+1,j}|^2=0,
\]
and by (\ref{EQ2.3}) we may conclude that
\begin{equation}
\label{EQ2.17}
c_{n+1,n+1} = a_{n+1,n+1},
\end{equation}
which upon substitution into (\ref{EQ2.12}) yields
\begin{equation}
\label{EQ2.18}
c_{n+1,j}= a_{n+1,j} \ \mbox{and} \
 c_{j,n+1}= a_{j,n+1} \ \mbox{for} \
n+2 \leq j \leq n^2,
\end{equation}
and together with (\ref{EQ2.9}) and (\ref{EQ2.11}) we conclude
\begin{equation}
\label{EQ2.19}
C[\{1,n+1\}|\{n+2,n+3, \ldots , n^2\}] =
A[\{1,n+1\}|\{n+2,n+3, \ldots, n^2\}],
\end{equation}
and
\begin{equation}
\label{EQ2.20}
C[\{n+2,n+3, \ldots , n^2\}|\{1,n+1\}]=
A[\{n+2,n+3, \ldots, n^2\}| \{1,n+1\}].
 \end{equation}

The analogue of (\ref{EQ2.4}) holds for $C$, and by (\ref{EQ2.18}),
(\ref{EQ2.19}) and $C[\{1,n+1\}]=A[\{1,n+1\}]$
we conclude
\[
C[\{n+2,n+3, \ldots , n^2\}]=
A[\{n+2,n+3, \ldots , n^2\}],
\]
leading finally to $C=A$, which in turn implies $D=A$.
Hence $A$ is an extreme point in this case.

\begin{itemize}
\item[(iib)] We can assume now that
\end{itemize}
\vspace{-10pt}
\[
a_{n+1,n+1}=a_{2n+1,2n+1}= \cdots = a_{(n-1)n+1,(n-1)n+1}=0,
\]
and this implies that rows and columns of $A$ indexed by $2,3,\ldots, n, n+1,2n+1, \ldots, (n-1)n+1$ are 0.
Define
\begin{equation}
\label{EQ2.21}
S = [n^2]\setminus \{1,2,\ldots, n,n+1,2n+1,\ldots, (n-1)n+1\} \ , \
S_1 = S \cup \{1\},
\end{equation}
and
$B=A[S]$,
$v = A[\{1\} | S ]$,
$\tilde{B} = A[S_1]$. So
\[
\tilde{B} = \left[\begin{array}{lc}
a_{11}&v\\v^*&B
\end{array}\right] \in
{\mathbb C}^{n^2-2n+2 \times n^2-2n+2}.
\]
Note that $B \in {\cal C}_{n-1}$, and its rank is 1 or 2. We distinguish 2 subcases.
\begin{itemize}
\item[(iib1)] Suppose that $B$ is an extreme point of ${\cal C}_{n-1}$.
We will show that $\dim {\cal F}(A)$ must be 0 or 2.
If $A$ is an extreme point we are done, so assume this is not the case.
\end{itemize}
Then there exist $F,G$ in ${\cal C}_n$, $F \neq G$,
such that $A = \frac{1}{2}(F+G)$.

Then, for $\Delta := A-F = (\delta_{ij}) \in {\mathbb C}^{n^2 \times n^2}$
we have
\[
G = A +\Delta \ , \
F = A - \Delta \ , \
B = \frac{1}{2}(F[S]+G[S]).
\]
Clearly, $F[S]$, $G[S] \in {\cal C}_{n-1}$,
and as $B$ is an extreme point of ${\cal C}_{n-1}$ (by assumption),
we must have
\begin{equation}
\label{EQ2.22}
B = F[S]=G[S] \Rightarrow
\Delta [S] =0.
\end{equation}
Let $w = \Delta[\{1\} | S]$. Then for $0 \leq t \leq 1$,
we have
\[
(A+t{\Delta})[S_1]=\left[\begin{array}{lc}
a_{11}&v+tw\\
v^*+tw^*&B
\end{array}\right].
\]
Moreover, since $\Delta \neq 0$ while $\Delta[S]=0$
and $\delta_{11}=0$
(as $b_{11}=g_{11}=a_{11}=1$),
it follows that $w \neq 0$. By symmetry we may assume
\begin{equation}
\label{EQ2.23}
\delta_{1,n+2}\neq 0.
\end{equation}
For $-1 \leq t \leq 1$, $A+t \Delta \in {\cal F}(A)$,
so its rank is at most 2.
Therefore, for $n+3 \leq j \leq n^2$, $j \not\equiv 1(\mbox{mod} \ n )$,
\[
\det \left[\begin{array}{ccc}
1&a_{1,n+2}+t\delta_{1,n+2}&a_{1j}+t\delta_{1j}\\
\bar{a}_{1,n+2}+t\bar{\delta}_{1,n+2} & a_{n+2,n+2}& a_{n+2,j}\\
\bar{a}_{1j}+t\bar{\delta}_{1j} &  \bar{a}_{n+2,j}&a_{jj}
\end{array}
\right]=0,
\]
and as $t \in [-1,1]$ is arbitrary, the coefficient of $t^2$ in the expansion must vanish, so
\begin{equation}
\label{EQ2.24}
-|\delta_{1,n+2}|^2a_{jj}-
|\delta_{1j}|^2a_{n+2,n+2}+
\delta_{1,n+2}\bar{\delta}_{1j}a_{n+2,j}+
\bar{\delta}_{1,n+2}\delta_{1j}\bar{a}_{n+2,j}=0.
\end{equation}
Hence, for the positive semidefinite matrix
$A[\{n+2,j\}]$ and for
$y:=\left[\begin{array}{c}
\delta_{1j}\\-\delta_{1,n+2}
\end{array}\right]$, we get from (\ref{EQ2.24}),
\[
y^*A[\{n+2,j\}]y =0,
\]
or, equivalently,
\begin{equation}
\label{EQ2.25}
A[\{n+2,j\}]y=0.
\end{equation}
Since $A$ is positive semidefinite it follows from
(\ref{EQ2.25}) that
\begin{equation}
\label{EQ2.26}
A^{(j)}=\frac{\delta_{1j}}{\delta_{1,n+2}}A^{(n+2)} \ \mbox{for} \
n+3 \leq j \leq n^2 \ , \ j \not\equiv 1 (\mbox{mod} \ n ).
\end{equation}
Note also that we must have
$\det A[\{1,n+2\}] > 0$, or else $A^{(1)}$ and $A^{(n+2)}$ are linearly dependent, leading to
$rank A =1$, a contradiction. In particular, we have
$a_{n+2,n+2}> 0$ and $|a_{1,n+2}|^2 < a_{n+2,n+2}$.

The previous discussion shows that $A^{(1)}$ and $A^{(n+2)}$ are linearly
independent, and the following is a basis for $\ker A$, and thus included
in the kernel of any matrix in ${\cal F}(A)$:
\[
\left\{e_2,e_3, \ldots , e_n,e_{n+1},e_{2n+1},\ldots, e_{(n-1)n+1}, \delta_{1,n+2}
e_j-\delta_{1j}e_{n+2} \ , \ j \in S \setminus \{n+2\}\right\}.
\]
Hence, given any $X =(x_{ij}) \in {\cal F}(A)$ we have $x_{11}=1$,
$X[[n^2]\setminus \{1\}]=A[[n^2]\setminus \{1\}]$ and
\[
X[\{1\} |S]=[x_{1,n+2}, \frac{\delta_{1,n+3}}{\delta_{1,n+2}}x_{1,n+2},
\frac{\delta_{1,n+4}}{\delta_{1,n+2}}x_{1,n+2}, \ldots , \frac{\delta_{1,n^2}}{\delta_{1,n+2}}x_{1,n+2}],
\]
with $|x_{1,n+2}|^2<a_{n+2,n+2}$. This shows that $\dim {\cal F}(A)=2$.

\vspace{10pt}

\noindent (iib2) It remains to consider the case that $B=A[S]$ is not an
extreme point of ${\cal C}_{n-1}$. This implies immediately that
$rank B = 2$, so there exist distinct $k,l \in S$ such that
$A^{(k)}$, $A^{(l)}$ are linearly independent, and
$\alpha_{k}, \alpha_{l} \in {\mathbb C}$ such that
\begin{equation}
\label{EQ2.27}
A^{(1)} = \alpha_k A^{(k)}+\alpha_l A^{(l)}.
\end{equation}

Define a linear map $\varphi:{\cal H}_{n^2} \rightarrow {\cal H}_{(n-1)^2}$
by $\varphi(H) = H[S]$.
We claim that $\varphi$ is 1-1 on ${\cal F}(A)$.
Indeed, suppose that $X,Y \in {\cal F}(A)$ with
$\varphi (X) = \varphi(Y)$.
It follows from (\ref{EQ2.27}) and Lemma \ref{lemma1.1} that
$e_1-\alpha_ke_k-\alpha_le_l \in ker A \subset \ker X,\ker Y$.
Hence $X^{(1)}=\alpha_kX^{(k)}+\alpha_lX^{(l)}$ and
$Y^{(1)}=\alpha_kY^{(k)}+\alpha_lY^{(l)}$, and since
$X[S]=Y[S]$ and $x_{11}=y_{11}=1$ it follows that $X=Y$.

\vspace{10pt}

By the induction hypothesis $\dim {\cal F}(B) \leq 2$ in ${\cal C}_{n-1}$
(actually we know it is positive by assumption that $B$ is not extreme),
and therefore in this subcase we can also conclude that
$\dim {\cal F}(A) \leq 2$, and combining with the previous subcases,
the same inequality holds always when $rank A=2$.

In the rest of the proof we show that all possible dimensions of
${\cal F}(L)$ are attained. This has already been demonstrated when
$n=2$ in the proof of (i), so assume now $n \geq 3$.

The following example shows that case (iia) can occur.\\ Let
$y=[e_1^t, 0^t,0^t, \ldots , 0^t]^t \in {\mathbb C}^{n^2}$
and $z=[0^t,e_1^t,e_2^t, \ldots, e^t_{n-1}]^t \in {\mathbb C}^{n^2}$,
and let $A=yy^*+zz^*$.
Clearly $A \in {\cal C}_n$ and $rank A =2$. Note that
$rank A[n+1,n+2, \ldots , n^2]=1$.

Next, we demonstrate that both possibilities that are discussed in (iib1)
can occur. Start with $B \in {\cal C}_{n-1}$ which is an extreme point and
with $rank B=2$, and we may assume without loss of generality that $b_{11}>0$.
To get an extreme point $A \in {\cal C}_n$ with $A[S]=B$
(where $S$ is defined by (\ref{EQ2.21})) let the rows and columns of $A$
indexed by $2,3,\ldots, n,n+1,2n+1, (n-1)n+1$ be 0. Let
$A[\{1\} | S] = \frac{1}{\sqrt{a_{n+2,n+2}}}A[\{n+2\} | S]=[b_{11}, b_{12},
\ldots , b_{1,(n-1)^2}]^t$,
$a_{11}=1$.
It follows that $A^{(1)}=\frac{1}{\sqrt{a_{n+2,n+2}}}A^{(n+2)}$, so
$e_1-\frac{1}{\sqrt{a_{n+2,n+2}}}e_{n+2}\in \ker A$,
hence to the kernel of every matrix in ${\cal F}(A)$.
This and the extremality of $B$ in ${\cal C}_{n-1}$ imply that $A$
is an extreme point of ${\cal C}_n$.

The second possibility arising in (iib1) is realized by starting with a
rank 1 extreme point $B \in {\cal C}_{n-1}$ (we may assume again $b_{11}>0$).
We define (hermitian) $A$ by letting $A[S]=B$, $a_{11}=1$, $A[\{1\} | S] =0$,
and all rows and columns of $A$ indexed by $2,3,\ldots , n , n+1, 2n+1, \ldots , (n-1)n+1$ be 0.
Then it is straightforward to see that $A \in {\cal C}_n$, and
$\dim {\cal F}(A) =2$.

Finally, we show that for any $n \geq 3$ there exists $A \in {\cal C}_n$
satisfying (iib2) and such that $\dim {\cal F}(A)=1$.
We start with $n=3$ and define
\begin{equation}
\label{EQ2.28}
P = \left[\begin{array}{ccc}
1&\frac{9}{4\sqrt{6}}&\frac{3}{2\sqrt{6}}\\
\frac{9}{4\sqrt{6}}&1&\frac{1}{4}\\
\frac{3}{2\sqrt{6}}&\frac{1}{4}&1
\end{array}\right].
\end{equation}
Then $P \in PSD_3$ and $Pq=0$, where $q=[1,-\frac{2}{\sqrt{6}},-\frac{1}{\sqrt{6}}]^t$.
Let
\begin{equation}
\label{EQ2.29}
A_3 = \left[\begin{array}{rrr}
E_{11}& \frac{9}{4\sqrt{6}}E_{12}&\frac{3}{2\sqrt{6}}E_{13}\\
 \frac{9}{4\sqrt{6}}E_{21}&E_{22} &\frac{1}{4}E_{23}\\
\frac{3}{2\sqrt{6}}E_{31}&\frac{1}{4}E_{32}& E_{33}
\end{array}\right]\in {\mathbb C}^{9 \times 9}.
\end{equation}
Then $A_3[\{159\}]=P$,
so $A_3 \in {\cal C}_3$ and $rank A_3=2$.
Note also $L(E_{11})=E_{11}$ (where $A_3=Z(L)$).
We claim that $\dim {\cal F}(A_3)=1$. The vectors $e_2,e_3,e_4,e_6,e_7,e_8$
and $e_1-\frac{2}{\sqrt{6}}e_5-\frac{1}{\sqrt{6}}e_9$ form a basis of
$\ker A_3$, thus belonging to the kernel of every matrix in ${\cal F}(A_3)$.
Let $X$ be any extreme point of ${\cal F}(A_3)$, then, necessarily,
$rank X =1$.
Also, $X^{(j)}=0$ for $j=2,3,4,6,7,8$.
There exist $x_1,x_2,x_3 \in {\mathbb C}$ such that
\[
X[\{1,5,9\}]=\left[\begin{array}{c}x_1\\x_2\\x_3\end{array}\right]
[\bar{x}_1 \ \bar{x}_2 \ \bar{x}_3],
\]
and we must also have $|x_i|=1$, $i=1,2,3$.
Without loss of generality assume $x_1=1$, and let $x_2=e^{-i\theta}$,
$x_3=e^{-i\varphi}$, where $\theta , \varphi \in {\mathbb R}$.
The condition $e_1-\frac{2}{\sqrt{6}}e_5-\frac{1}{\sqrt{6}}e_9 \in \ker X$
leads to the equation (over $\mathbb C$)
\[
2e^{i\theta}+e^{i\varphi}=\sqrt{6},
\]
or, equivalently, two equations (over $\mathbb R$)
\begin{equation}
\label{EQ2.30}
2\cos \theta +\cos \varphi = \sqrt{6}, \ \mbox{and} \
2\sin \theta +\sin \varphi =0.
\end{equation}

Since $\sqrt{6} > 2$ it follows from (\ref{EQ2.30}) that $\cos \theta, \cos \varphi > 0$, so
$\theta , \varphi \in (-\frac{\pi}{2},\frac{\pi}{2})$.
Moreover, $\sin \theta$ and $\sin \varphi$ must have opposite signs.
We have
\[
1-4\sin^2 \theta = 1-\sin^2 \varphi = \cos^2 \varphi = 6-4\sqrt{6}\cos \theta +4\cos^2 \theta,
\]
so
\[
4\sqrt{6}\cos \theta =9,
\]
implying
\begin{equation}
\label{EQ2.31}
\cos \theta = \frac{9}{4\sqrt{6}} \ \mbox{and} \
\sin \theta = \pm \frac{1}{4}\sqrt{\frac{5}{2}}.
\end{equation}
Hence
\begin{equation}
\label{EQ2.32}
\cos \varphi = \sqrt{6}-\frac{9}{2\sqrt{6}}=\frac{3}{2\sqrt{6}}
\ \mbox{and} \
\sin \varphi = \mp \frac{1}{2} \sqrt{\frac{5}{2}}.
\end{equation}
Thus, (\ref{EQ2.30}) has 2 solutions, $(\theta_1, \varphi_1)$ with
$0 < \theta_1< \frac{\pi}{2}$, $-\frac{\pi}{2}< \varphi_1 < 0$,
and $(\theta_2, \varphi_2)$ with
$-\frac{\pi}{2} < \theta_2<0$, $0 < \varphi_2 < \frac{\pi}{2}$,
corresponding to the expressions obtained in (\ref{EQ2.31}) and (\ref{EQ2.32}),
including the signs in the appropriate order.
This shows that ${\cal F}(A_3)$ has 2 extreme points, $X$ and $X_1$.
Here we write only their nontrivial part, namely $X[\{159\}]$ and
$X_1[\{159\}]$.
\[
X[\{159\}] =\left[\begin{array}{ccc}
1& e^{i\theta_1}& e^{i\varphi_1}\\
e^{-i\theta_1}& 1 & e^{i(\varphi_1-\theta_1)}\\
e^{-i\varphi_1}& e^{i(\theta_1-\varphi_1)}&1
\end{array}\right] \ , \
X_1[\{159\}] =\left[\begin{array}{ccc}
1& e^{i\theta_2}& e^{i\varphi_2}\\
e^{-i\theta_2}& 1 & e^{i(\varphi_2-\theta_2)}\\
e^{-i\varphi_2}& e^{i(\theta_2-\varphi_2)}&1
\end{array}\right] ,
\]
so
\[
X[\{159\}] =\left[\begin{array}{ccc}
1&\frac{9}{4\sqrt{6}}+\frac{\sqrt{5}}{4\sqrt{2}}i & \frac{3}{2\sqrt{6}}-\frac{\sqrt{5}}{2\sqrt{2}}i\\
\frac{9}{4\sqrt{6}}-\frac{\sqrt{5}}{4\sqrt{2}}i& 1& \frac{1}{4} -\frac{\sqrt{15}}{4}i\\
\frac{3}{2\sqrt{6}}+\frac{\sqrt{5}}{2\sqrt{2}}i& \frac{1}{4} +\frac{\sqrt{15}}{4}i&1
\end{array}\right]= \bar{X}_1[\{159\}].
\]
Note that $A = \frac{1}{2}(X+X_1)$.

\vspace{10pt}

The construction of $A \in {\cal C}_n$ satisfying (iib2) and
$\dim {\cal F}(A)=1$ is based on the construction of $A_3$ in (\ref{EQ2.29}).
Let $n \geq 3$ and $A^{[n]}=[A_{ij}]_{i,j =1,2,\ldots , n} \in {\mathbb C}^{n^2 \times n^2}$,
where $A_{ij}=E_{ij}\in {\mathbb C}^{n \times n}$ for $i.j \in [n-2]$;
$A_{i,n-1}= A^t_{n-1,i}=\frac{9}{4\sqrt{6}}E_{i,n-1} \in {\mathbb C}^{n \times n}$ and
$A_{in}=A^t_{ni}=\frac{3}{2\sqrt{6}}E_{in} \in {\mathbb C}^{n \times n}$ for $i \in [n-2]$;
$A_{ii}=E_{ii} \in {\mathbb C}^{n \times n}$ for $i=n-1,n$;
$A_{n-1,n}=A^t_{n,n-1}=\frac{1}{4}E_{n-1,n} \in {\mathbb C}^{n \times n}$.

Note that $A^{[3]}=A_3$, and clearly $A^{[n]}\in {\cal C}_n$.
Moreover, for the corresponding quantum channel $L$
(so that $A^{[n]} =Z(L)$) we have $L(E_{11})=E_{11}$.
Furthermore, whenever $n \geq 4$, for the set $S$ defined in
(\ref{EQ2.21}), we have $A^{[n]}[S] =A^{[n-1]}$,
so $A^{[n-1]}$ is embedded as a principal submatrix in $A^{[n]}$.

\vspace{10pt}

Let $S_2=\{(n+1)i+1, i=0,1,2,\ldots,n-1\}$.
Then the only nonzero rows (and columns) of $A^{[n]}$
are those indexed by elements in $S_2$, and we have
\[
A^{[n]}[S_2]=\left[\begin{array}{rcc}
J_{n-2}&\frac{9}{4\sqrt{6}}J_{n-2,1}&
\frac{3}{2\sqrt{6}}J_{n-2,1}\\
\frac{9}{4\sqrt{6}}J_{1,n-2}&1 & \frac{1}{4}\\
\frac{3}{2\sqrt{6}}J_{1,n-2}& \frac{1}{4}&1
\end{array}\right].
\]

The latter matrix is congruent to the direct sum of $P$,
defined in (\ref{EQ2.28}), and the 0 matrix of order $n-3$,
hence $rank\ A^{[n]}=2$.
Therefore any extreme matrix $X$ in ${\cal F}(A^{[n]})$ must have $rank\ 1$,
and its kernel must contain $\ker A^{[n]}$. Write $X=xx^*$,
$x=(x_j) \in {\mathbb C}^{n^2}$. Then $x_j =0$ for $j \not\in S_2$, and
$|x_j|=1$ for $j \in S_2$.
We may normalize so that $x_1=1$, and $\ker X$ forces $x_j =1$ for any $j \in S_2$,
except for $j=(n+1)(n-2)+1$ and $j=(n+1)(n-1)+1=n^2$. The proof proceeds now as in the case of $n=3$.
\hspace{\fill}$\Box$
\section{Proper faces of ${\cal L}_n$ of maximum dimension}
\setcounter{equation}{0}

Our main goal here is to compute the maximum dimension of a proper face of
${\cal L}_n$, or equivalently ${\cal C}_n$.
We start with the following lemma.

\begin{lemma}
\label{lemma3.1}
Let $A \in {\cal C}_n$ with $r = rank A$, and assume $r \leq n^2-2$.
Then ${\cal F}(A)$ is strictly contained in a proper face of
${\cal C}_n$ generated by a matrix of rank $r+1$.
\end{lemma}

\noindent {\bf Proof:} There exists $x=[x_1^t, x_2^t, \ldots , x_n^t]^t \in {\mathbb C}^{n^2} \neq 0$,
$x \in \ker A$, where
$x_i = (x_i^{(j)}) \in {\mathbb C}^n$, $i \in [n]$.
Therefore, there exist $1 \leq i,j \leq n$ such that
$x_i^{(j)}\neq 0 $.
Let $\sigma$ be a permutation of $[n]$ such that
$\sigma (i) =j$,
and let $u = [e^t_{\sigma(1)}, e^t_{\sigma(2)}, \ldots , e^t_{\sigma(i-1)}, e^{i\theta}e^t_{\sigma(i)}, \ldots , e^t_{\sigma(n)}]^t \in {\mathbb C}^{n^2}$,
with $\theta \in {\mathbb R}$ to be determined.
By  Observation (\ref{observe1.1}) $uu^* \in {\cal C}_n $,
and $\theta$ can be chosen so that $\langle u, x \rangle \neq 0$.
Let $A_1 = \frac{1}{2}(A+uu^*)$. Then
$A_1 \in {\cal C}_n$ with $rank A_1 = r+1 < n^2$, and
${\cal F}(A_1)$ is a proper face of ${\cal C}_n$, strictly containing ${\cal F}(A)$. \hspace{\fill} $\Box$

It follows from the lemma that to compute the maximum dimension of a proper face of ${\cal C}_n$ it suffices to consider matrices in ${\cal C}_n$ of rank $n^2-1$.

\begin{observation}
\label{observe3.1}
Suppose that $L \in {\cal L}_n$ with
$A =Z[L]=[A_{ij}]_{i,j=1}^n$,
and suppose that $z=[z_1^t, z_2^t, \ldots , z_n^t]^t\in \ker A$,
with $z_i \in {\mathbb C}^n$, $i \in [n]$.
Let $U \in {\mathbb C}^{n \times n}$ be any unitary matrix.
Then $A_1 = [UA_{ij}U^*]_{i,j=1}^{n}=Z[L_1]$ for some $L_1 \in {\cal L}_n$,
and $[(Uz_1)^t, (Uz_2)^t , \ldots , (Uz_n)^t]^t \in \ker A_1$.
\end{observation}

\noindent {\bf Proof:} Since $A \in {\cal C}_n$ so is $A_1$, and the rest is clear. \hspace{\fill} $\Box$

\begin{theorem}
\label{thm3.1}
Let $n \geq 2$ be an integer. Then the maximum dimension of a proper face of
${\cal L}_n$ (equivalently ${\cal C}_n$) is $n^4-3n^2+1$.
\end{theorem}

\noindent {\bf Proof:} We consider ${\cal C}_n$, and by Lemma \ref{lemma3.1}
it suffices to consider faces generated by matrices in ${\cal C}_n$ of rank
$n^2-1$. We start with $n=2$, which has to be dealt with separately.

\vspace{10pt}

\noindent \underline{$n=2$} Suppose that $A \in {\cal C}_2$ with $rank A =3$,
and  let $z =[v^t, w^t]^t$, $v,w \in {\mathbb C}^2$ be a nonzero vector in
$\ker A$ (it is uniquely determined up to a scalar multiple). We may assume
without loss of generality that $v \neq 0$, and using scaling and Observation
\ref{observe3.1} we may assume $v=e_1$.

Suppose first that $v,w$ are linearly dependent, so
$w = \alpha e_1$ for some $\alpha \in {\mathbb C}$. Then $Az=0$ yields the following system:
\[
\begin{array}{r}
a_{11}=-\alpha a_{13},\\
\bar{a}_{12}=-\alpha a_{23},\\
\bar{a}_{13}=-\alpha a_{33},\\
\bar{a}_{14}=-\alpha \bar{a}_{34},\\
\end{array}
\]
which together with the trace conditions imposed by $A \in {\cal C}_2$ yield
\[
A =\left[\begin{array}{cccc}
|\alpha|^2a_{33}& -\bar{\alpha}\bar{a}_{23}& -\bar{\alpha}{a}_{33}& -\bar{\alpha}{a}_{34}\\
-{\alpha}{a}_{23}&1-|\alpha|^2a_{33}& a_{23}&  \bar{\alpha}{a}_{33}\\
-{\alpha}{a}_{33}& \bar{a}_{23}&a_{33}& a_{34}\\
-\alpha \bar{a}_{34}& \alpha a_{33}& \bar{a}_{34}& 1-a_{33}
\end{array}\right].
\]
As $a_{33}$, $a_{23}$ and $a_{34}$ are free variables (subject to
$A \in PSD_4$), we see that $\dim {\cal F}(A)=5$.

We may assume now that $v,w$ are linearly independent, so let
$w=[\alpha , \beta]^t \in {\mathbb C}^2$, with $\beta \neq 0$.
Then $Az=0$ yields the following system:

\[
\begin{array}{l}
a_{11}+ \alpha a_{13}+ \beta a_{14}=0,\\
\bar{a}_{12}+ \alpha a_{23}- \beta a_{13}=0,\\
\bar{a}_{13}+\alpha a_{33}+\beta a_{34}=0,\\
\bar{a}_{14}+ \alpha \bar{a}_{34}+\beta(1-a_{33})=0.
\end{array}
\]
Note that we must have $0 < a_{11}< 1$ and  $0< a_{33}< 1$,
or else $rank A=2$. It follows from the system that
\[
\begin{array}{rl}
a_{13}&=-\bar{\alpha}a_{33}-\bar{\beta}\bar{a}_{34},\\
a_{14}&= -\bar{\alpha}a_{34} - \bar{\beta}(1-a_{33}),\\
a_{12}&= -\bar{\alpha}\bar{a}_{23}+\bar{\beta}(-\alpha a_{33}-\beta a_{34})= -\bar{\alpha}\bar{a}_{23}-\alpha \bar{\beta}a_{33}-|\beta|^2a_{34},\\
a_{11}&= -\alpha (-\bar{\alpha}a_{33}-\bar{\beta}\bar{a}_{34})
 -\beta (-\bar{\alpha}a_{34}-\bar{\beta}+ \bar{\beta}a_{33})=\\
 &= (|\alpha|^2-|\beta|^2)a_{33}+\alpha \bar{\beta}\bar{a}_{34}+\bar{\alpha}\beta a_{34}+|\beta|^2,
\end{array}
\]
which together with $\gamma := |\alpha|^2-|\beta|^2$ and the trace conditions yield
\[
A =\]

\[\footnotesize
\left[\begin{array}{cccc}
\gamma a_{33}+2\mbox{Re} (\alpha \bar{\beta}\bar{a}_{34}) +|\beta|^2&
-\bar{\alpha}\bar{a}_{23}-\alpha\bar{\beta}a_{33}-|\beta|^2a_{34}&
-\bar{\alpha}a_{33}-\bar{\beta}\bar{a}_{34}&
-\bar{\alpha}a_{34}-\bar{\beta}(1-a_{33})\\
-{\alpha}{a}_{23}-\bar{\alpha}{\beta}a_{33}-|\beta|^2\bar{a}_{34}&
1-\gamma a_{33}-2\mbox{Re} (\alpha \bar{\beta}\bar{a}_{34}) -|\beta|^2&
a_{23}& \bar{\alpha}a_{33}+\bar{\beta}\bar{a}_{34}\\
-{\alpha}a_{33}-{\beta}{a}_{34}& \bar{a}_{23}& a_{33}&a_{34}\\
-{\alpha}\bar{a}_{34}-{\beta}(1-a_{33})& \alpha a_{33}+ \beta a_{34}&
\bar{a}_{34}& 1-a_{33}
\end{array}\right].
\]
As $a_{33}$, $a_{23}$ and $a_{34}$ are free variables (subject to $A \in PSD_4$), we get
$\dim {\cal F}(A)=5$.

\vspace{10pt}

\noindent \underline{$n\geq 3$} Suppose that $A \in {\cal C}_n$ such that
$rank A = n^2-1$. Then, there exists $x = (x_i) \in {\mathbb C}^{n^2}$
(unique up to scalar multiples) such that $Ax=0$.
We may assume without loss of generality, using scaling and Observation \ref{observe3.1}, that $x_{n^2}=1$ and $x_{n^2-1}\neq 0$.

Write $A =[A_{ij}]_{i,j=1}^{n}$, with $A_{ij}\in {\mathbb C}^{n \times n}$.
By assumption, $A$ has to satisfy the linear equations
$tr A_{ij}=\delta_{ij}$ for
$i,j \in [n]$, and $Ax=0$.
We will show that $\dim {\cal F}(A)=n^4-3n^2+1$ by using these equations to
express certain entries of $A$ in terms of the other entries.
More precisely, the following discussion will show that the following are
not free variables:
$(i) a_{i,n^2}$, $i \in [n^2]$ and $i \not\equiv 0(\mbox{mod} \ n)$;
$a_{kn,n^2-1}$, $k \in [n-1]$;
$a_{n^2,n^2}$;
For $1 \leq i\leq j \leq n$, the first entry on the main diagonal of
$A_{ij}$.

We use first the trace conditions. For $1 \leq i \leq j \leq n$ we use
$tr A_{ij}=\delta_{ij}$ to express the first entry on its main
diagonal in terms of its successors. So, for example,
$a_{11}=1-\sum\limits_{i=2}^{n}a_{ii}$, $a_{1,n+1}=
- \sum\limits_{i=2}^{n}a_{i,n+i}$, etc.

Next we consider the kernel condition, namely the
$n^2$ homogeneous, linear equations $(Ax)_i=0$,
$i \in [n^2]$.
We consider them sequentially, and as we will see, eliminating at each
step exactly one of the remaining non-free variables that is, either
$a_{i,n^2}$ or $a_{i,n^2-1}$ for a suitable $i$. We consider these equations
in the natural order.

\vspace{10pt}

Suppose that $i \in [n^2-1]$.
When $i \not\equiv 0 (\mbox{mod}\ n)$ we use $(Ax)_i=0$ to eliminate
$a_{i,n^2}$.
Let $f_i^{(n^2)}$ be the (linear) expression obtained when writing
$a_{i,n^2}$ in terms of free variables.
When $i \equiv 0 (\mbox{mod}\ n)$, $i < n$, we use $(Ax)_i=0$ to eliminate
$a_{i, n^2-1}$.
Let $f_i^{(n^2-1)}$ be the (linear) expression obtained when writing
$a_{i,n^2-1}$
in terms of free variables (this is possible since $x_{n^2-1}\neq 0$).

For example,
\[
\begin{array}{l}
f_1^{(n^2)}= -\sum\limits_{j=1}^{n^2-1}a_{1j}x_j=-\sum\limits_{\begin{array}{c}
\mbox{\scriptsize$j=1$} \\\mbox{\scriptsize $j \not\equiv 1 (\mbox{mod}\ n)$}
\end{array}}^{n^2-1}a_{1j}x_j-1x_1+\sum\limits_{i=2}^{n}\sum\limits_{k=0}^{n-1}a_{i,i+kn} x_{1+kn} , \\
f_n^{(n^2-1)} = -\frac{1}{x_{n^2-1}}\cdot
\sum\limits_{\begin{array}{c}
\mbox{\scriptsize$j=1$}\\\mbox{\scriptsize$j \neq n^2-1$}
\end{array}}^{n^2}
a_{nj}x_j=
-\frac{1}{x_{n^2-1}}\left( \sum\limits_{j=1}^{n-1}\bar{a}_{jn}x_j+ \sum\limits_{\begin{array}{c}
\mbox{\scriptsize$j=n$}\\\mbox{\scriptsize$j \neq n^2-1$}
\end{array}}^{n^2}a_{nj}x_j\right).
\end{array}
\]
We finally get to the last equation, that is, $(Ax)_{n^2}=0$, and show it will determine $a_{n^2,n^2}$.
Note that $a_{n^2,n^2}$ appears also in the equation $(Ax)_{(n-1)n+1}=0$,
which has already been discussed.
Hence it appears in $f_{n^2-n+1}^{(n^2)}$, and so it appears in the summand
$a_{n^2,n^2-n+1}x_{n^2-n+1}=\bar{f}_{n^2-n+1}^{(n^2)}x_{n^2-n+1}$ in
$(Ax)_{n^2}=0$. So
$a_{n^2,n^2}$ appears twice in the last equation. Recall that $a_{n^2,n^2}$
is real, and we will prove that the last equation determines $a_{n^2,n^2}$
in a well defined way, completing the counting of the non-free variables.

\vspace{10pt}

Consider  now the appearance of each free variable in the last equation.
We distinguish several cases, each dealing with a free variable.
As $A$ is hermitian, we have $a_{ji}=\bar{a}_{ij}$, so free variables
appear on the main diagonal or above. The first two cases deal with main
diagonal entries.
\begin{itemize}
\item[({\rm I})] Consider a main diagonal entry $a_{ii}$, where
$i \not\equiv 1, n (\mbox{mod} \ n)$.
Write $i =qn+r$, where $0 \leq q \leq n-1$ and
$2 \leq r \leq n-1$. Then $a_{ii}$ appears in $(Ax)_i=0$ and
$(Ax)_{i-r+1}=0$,
hence in $f_i^{(n^2)}$ and $f_{i-r+1}^{n^2}$ with coefficients
$-x_i$ and $x_{i-r+1}$, respectively.
Therefore, the summands in the left hand side of $(Ax)_{n^2}=0$ containing
$a_{ii}$ are $-\overline{a_{ii}x_i}x_i+\overline{a_{ii}x_{i-r+1}}x_{i-r+1}a_{ii}(|x_{i-r+1}|^2-|x_i|^2)\in {\mathbb R}$.
\item[({\rm II})] Consider a main diagonal entry $a_{ii}$, where
$i\equiv 0(\mbox{mod}\ n)$, $i\leq n^2-1$.
Then $a_{ii}$ appears in $(Ax)_i=0$ and $(Ax)_{i-n+1}=0$. The latter yields
that $a_{ii}$ appears in $f_{i-n+1}^{(n^2)}$ with coefficient $x_{i-n+1}$.
The former yields that $a_{ii}$ appears in $f_i^{(n^2-1)}$ with coefficient
$-\frac{x_i}{x_{n^2-1}}$.
Hence $a_{ii}$ appears in $(Ax)_{n^2-1}=0$, and so it appears in
$f^{(n^2)}_{n^2-1}$ with coefficient $\frac{|x_i|^2}{\bar{x}_{n^2-1}}$. Hence, the summands in the left hand side of $(Ax)_{n^2}=0$ containing $a_{ii}$ are
$a_{ii}(|x_{i-n+1}|^2+|x_i|^2) \in {\mathbb R}$.
\item[({\rm III})] Consider $a_{ij}$ (a free variable) that does not appear on
either of the main diagonal, last row, last column of the block $A_{kl}$
containing it. Then $a_{ij}$ appears in $f_i^{(n^2)}$ with coefficient
$-x_j$ and $a_{ji}=\bar{a}_{ij}$ appears in $f_j^{(n^2)}$ with coefficient $-x_i$.
Hence, the summands in the left hand side of $(Ax)_{n^2}=0$ containing
$a_{ij}, a_{ji}$ are
$-\overline{a_{ij}x_j}x_i - \overline{a_{ji}x_i}x_j=-2\mbox{Re}(a_{ij}x_j\bar{x}_i) \in {\mathbb R}$.
\item[({\rm IV})] Consider $a_{ij}$ (a free variable) that appears in the
last column of the block $A_{kl}$ containing it, but not on its main diagonal.
So $i \not\equiv 0 (\mbox{mod}\ n)$, $j \equiv 0 (\mbox{mod}\ n)$. Then
$a_{ij}$ appears in $f_i^{(n^2)}$ with coefficient $-x_j$, and $a_{ji}=\bar{a}_{ij}$ appears in $f_j^{(n^2-1)}$ with coefficient $-\frac{x_i}{x_{n^2-1}}$.
This implies that $\bar{a}_{ji}$ appears in $f_{n^2-1}^{(n^2)}$ with
coefficient $\frac{\bar{x}_ix_j}{\bar{x}_{n^2-1}}$.
Hence, the summands in the left hand side of $(Ax)_{n^2}=0$ containing
$a_{ij}, a_{ji}$ are
$-\overline{a_{ij}x_j}x_i+ a_{ji}x_i\bar{x}_j=0$.
\item[({\rm V})] Consider $a_{ij}$ ( a free variable) that appears in the last row of the block $A_{kl}$ containing it, but not on its main diagonal.
So $i \equiv 0 (\mbox{mod}\ n)$ and $j \not\equiv 0 (\mbox{mod}\ n)$.
Then $a_{ij}$ appears in $f_i^{(n^2-1)}$ with coefficient
$-\frac{x_j}{x_{n^2-1}}$, so  $\bar{a}_{ij}={a}_{ji}$ appears in
$f_{n^2-1}^{(n^2)}$ with coefficient $\frac{\bar{x}_jx_i}{\bar{x}_{n^2-1}}$.
Also,  $a_{ji}=\bar{a}_{ij}$ appears in $f_{j}^{(n^2)}$ with
coefficient $-x_i$.
Hence, the summands in the left hand side of $(Ax)_{n^2}=0$ containing
$a_{ij}, a_{ji}$ are
$a_{ij}x_j\bar{x}_i- a_{ij}\bar{x}_i{x}_j=0$.
 \item[({\rm VI})] Consider  a free variable $ a_{i,n^2}$,
so necessarily $i \equiv 0 (\mbox{mod}\ n)$ and $i < n^2$.
Then $ a_{i,n^2}$ appears in $f_i^{(n^2-1)}$ with coefficient
$-\frac{1}{x_{n^2-1}}$, so  $\bar{a}_{i,n^2}$ appears in
$f_{n^2-1}^{(n^2)}$ with coefficient $\frac{{x}_i}{\bar{x}_{n^2-1}}$.
Also, $a_{i,n^2}$ appears  as a summand in the expression for
$a_{i-n+1,n^2-n+1}$, and its coefficient in $f_{i-n+1}^{(n^2)}$ is $x_{n^2-n+1}$.
On the other hand $a_{n^2,i}=\bar{a}_{i,n^2}$ appears in the last equation
with coefficient $x_i$, and also in $f_{n^2-n+1}^{(n^2)}$ with coefficient $x_{i-n+1}$. Hence, the summands in the left hand side of $(Ax)_{n^2}=0$ containing
$a_{i,n^2}, a_{n^2,i}$ are

$a_{i,n^2}\bar{x}_i+\bar{a}_{i,n^2}\bar{x}_{n^2-n+1}x_{i-n+1}+
\bar{a}_{i,n^2}x_i+ {a}_{i,n^2}\bar{x}_{i-n+1}x_{n^2-n+1} \in {\mathbb R}$
\end{itemize}
 The next two cases deal with a free variable that lies on the main diagonal of the block containing it, but not on the last column of $A$.
\begin{itemize}
\item[({\rm VII})]  Consider  a free variable $a_{ij}$, such that
 $i \not\equiv 0 (\mbox{mod}\ n)$ and $j \equiv i (\mbox{mod}\ n)$ and $j < n^2$.
Write $i=qn+r$, where $0 \leq q \leq n-1$ and $1<r \leq n-1$
(note that we cannot have $r=1$).
Then $a_{ij}$ appears in $f_i^{(n^2)}$ with coefficient
$-x_j$, and it also appears as a summand in the expression for
$a_{i-r+1,j-r+1}$.
Hence it appears in
$f_{i-r-1}^{(n^2)}$ with coefficient $x_{j-r+1}$.
Also,  $a_{ji}=\bar{a}_{ij}$ appears in $f_{j}^{(n^2)}$ with
coefficient $-x_i$ and in
$f_{j-r+1}^{(n^2)}$ with
coefficient $x_{i-r+1}$.
Hence, the summands in the left hand side of $(Ax)_{n^2}=0$ containing
$a_{ij}, a_{ji}$ are
$-\bar{a}_{ij}\bar{x}_j{x}_i+ \bar{a}_{ij}\bar{x}_{j-r+1}{x}_{i-r+1}-a_{ij}\bar{x}_ix_j+ a_{ij}\bar{x}_{i-r+1}x_{j-r+1} \in {\mathbb R}$.
\item[({\rm VIII})] Consider  a free variable $a_{ij}$, such that
 $i \equiv 0 (\mbox{mod}\ n)$ and $j \equiv i (\mbox{mod}\ n)$ and $j < n^2$.
Then $a_{ij}$ appears in $f_i^{(n^2-1)}$ with coefficient
$-\frac{x_j}{x_{n^2-1}}$, so $\bar{a}_{ij}={a}_{ji}$ appears in
$f_{n^2-1}^{(n^2)}$ with
coefficient $\frac{\bar{x}_jx_i}{{x}_{n^2-1}}$.
Also,  $a_{ij}$ is a summand
in
$a_{i-n+1,j-n+1}$ with  coefficient $-x_{j-n+1}$, so it
 appears in $f_{i-n+1}^{(n^2)}$ with
coefficient $x_{j-n+1}$.
In addition, $a_{ji}= \bar{a}_{ij}$
appears in
$f_{j}^{(n^2-1)}$ with
coefficient $-\frac{x_i}{x_{n^2-1}}$, so $a_{ij}$
appears in
$f_{n^2-1}^{(n^2)}$ with
coefficient $\frac{\bar{x}_ix_j}{\bar{x}_{n^2-1}}$.
Also,  $a_{ji}$
appears in
$f_{j-n+1}^{(n^2)}$ with
coefficient $x_{i-n+1}$.
Hence, the summands in the left hand side of $(Ax)_{n^2}=0$ containing
$a_{ij}, a_{ji}$ are
$a_{ij}x_j\bar{x}_i+\bar{a}_{ij}\bar{x}_{j-n+1}x_{i-n+1}+
\bar{a}_{ij}x_i\bar{x}_j+a_{ij}\bar{x}_{i-n+1}x_{j-n+1}\in {\mathbb R}$.
\item[({\rm IX})] Finally we consider the coefficient of $a_{n^2,n^2}$
in the last equation. As $a_{n^2,n^2}$ appears in a summand in the expression for
$a_{n^2-n+1,n^2-n+1}$ with coefficient $-x_{n^2-n+1}$, it appears in
$f_{n^2-n+1}^{(n^2)}$ with coefficient $x_{n^2-n+1}$, so the coefficient of $a_{n^2,n^2}$ in the last equation is $|x_{n^2-n+1}|^2+1 >0$.
As all other summands in the left hand side of $(Ax)_{n^2}=0$ are real,
as we have shown, it follows that this equation determines $a_{n^2,n^2}$,
completing the proof. \hspace{\fill} $\Box$
\end{itemize}

\end{document}